\documentstyle[amssymb,prl,aps,multicol,epsfig]{revtex}

\newcommand{\sm}{SmLa$_{0.8}$Sr$_{0.2}$CuO$_{4-\delta}$}

\begin{document}
\title{Magnetic Field Dependence of the Transverse Plasmon in
SmLa$_{0.8}$Sr$_{0.2}$CuO$_{4-\delta}$}
\author{A. Pimenov$^{1}$, A. Loidl$^{1}$, D. Duli\'c$^{2}$,
D. van der Marel$^{2}$, I. M. Sutjahja$^{3}$
and A. A. Menovsky$^{3}$}
\address{$^{1}$Experimentalphysik V, EKM, Universit\"{a}t Augsburg,
86135 Augsburg, Germany}
\address{$^{2}$Laboratory of Solid State
Physics, Materials Science Centre, 9747 AG Groningen, The
Netherlands}
\address{$^{3}$Van der Waals - Zeeman Institut, University of Amsterdam, The
Netherlands}
\date{July 30, 2001}
\maketitle

\begin{abstract}
The magnetic field and temperature dependence of the transverse
and longitudinal plasmons in SmLa$_{0.8}$Sr$_{0.2}$CuO$_{4-\delta
}$ have been measured for frequencies $5<\nu <30$\,cm$^{-1}$ and
magnetic fields $0<B<7$\,T. A transition between a vortex glass
and a vortex liquid regime, which revealed different field
dependencies of the resonance frequencies, could be clearly
observed. The positions and the spectral weights of the plasmons
were successfully described using the multilayer model (Phys. Rev.
B {\bf 64}, 024530, 2001), which takes the compressibility of the
electronic liquid into account. The compressibility is shown to
remain constant in the broad range of field and temperature,
which strongly supports the applicability of the model. The
absolute value of the compressibility is close to that of a
two-dimensional noninteracting electron gas.
\end{abstract}


\begin{multicols}{2}

The Josephson coupling of the CuO$_{2}$ layers in the
high-temperature superconductors leads to the appearance of a
 plasma resonance (JPR) along the c-axis \cite{tachiki}.
The frequency of this resonance is determined by the strength of
the Josephson coupling between the layers and is directly
connected to the c-axis penetration depth. The magnetic field and
temperature dependence of the JPR has been intensively
investigated during the last years providing important
informations on the c-axis transport and on the vortex dynamics
\cite{tsui,matsuda,shibaushi,gaifullin}. Recently, static stripe
ordering has been observed in
La$_{1.85-y}$Nd$_{y}$Sr$_{0.15}$CuO$_{4}$ \cite{tajima} using this
technique.

Additional plasma resonances have been predicted \cite{marel1} for
systems with two different coupling constants between the
CuO$_{2}$ layers. For such systems two longitudinal and a
transverse plasma modes are expected. The transverse excitation
couples directly to the electromagnetic radiation and can
therefore be observed as a peak in the real part of the optical
conductivity. To explain the appearance of transverse plasma
oscillations a simple model with alternating coupling constants
(multilayer model) has been proposed by van der Marel and
Tsvetkov \cite{marel1}.

The investigation of the transverse plasma has proved to be
difficult in YBaCuO \cite{ybco} and BiSrCaCuO \cite{bcco} because
the characteristic maxima overlap with the phonon resonances. In
contrast, the characteristic plasma frequencies of
SmLa$_{1-x}$Sr$_{x}$CuO$_{4-\delta }$ \cite{shibata1} are shifted
down to the submillimeter frequency range and therefore can well
be separated from the phonons. Recently, several groups have
reported the observation of two longitudinal and one transverse
plasmons in single crystalline SmLa$_{1-x}$Sr$_{x}$CuO$_{4-\delta
}$ \cite{shibata2,kakeshita,dulic} and
Nd$_{1.4}$Sr$_{0.4}$Ce$_{0.2}$CuO$_{4-\delta }$ \cite{kakeshita}.
However, detailed comparison between the experimental data and the
multilayer model predictions revealed substantial discrepancies
concerning the position and the spectral weight of the transverse
plasmon \cite{shibata2,kakeshita}. It has been proposed
\cite{marel2} that the electronic compressibility should be taken
into account to bring theory and experiment in better agreement.
And indeed, the extended version of the multilayer model including
electronic compressibility  \cite{marel2} was able to successfully
reproduce the position and amplitude of the transverse plasmon
\cite{dulic}.

In this paper we present the magnetic field and temperature
dependencies of the transverse and longitudinal plasmons in a
single crystal of SmLa$_{0.8}$Sr$_{0.2}$CuO$_{4-\delta }$
(SmLSCO). The extended version of the multilayer model
\cite{marel2} has been applied to describe the observed spectra.
We analyze the dependences of the model parameters with special
attention to the electronic compressibility, which in the first
approximation may be expected to be independent on magnetic field
and temperature.

The single crystals of SmLSCO were grown using a four-mirror image
furnace as described previously \cite{dulic}. After growth the
sample has been annealed in the oxygen under high pressure. The
crystal showed a superconducting transition at $T_{c}=16$\,K with
the transition width of 2 K. A thin ac-oriented plate of the size
$\sim 3\times 4$\,mm$^{2}$ has been cut from the original crystal
and polished to a thickness of $\delta \simeq 120$\,$\mu m$ by
diamond paste. The transmittance measurements in the
submillimeter-wave range have been performed using a coherent
source spectrometer \cite{kozlov}. Different backward-wave
oscillators (BWO's) have been employed as monochromatic and
continuously tunable sources covering the range from 5\,cm$^{-1}$
to 30\,cm$^{-1}$. The Mach-Zehnder interferometer arrangement has
allowed measuring both the intensity and the phase shift of the
wave transmitted through the sample. Using the Fresnel optical
formulas for the complex transmission coefficient of a
plane-parallel sample, the complex conductivity has been
determined directly from the measured spectra. Due to high degree
of the polarization of the radiation, the spectra could be, in
principle, measured both along the a- and c-axes. However, due to
high conductivity of SmLSCO within the CuO$ _{2}$ planes, the
sample was completely opaque for $\tilde{e}\parallel a$-axis
($\tilde{e}$ is the ac-electric field of the incident radiation).
The anisotropy could be measured solely for room temperature as
$\sigma _{a}/\sigma _{c}\simeq 19\pm 3$.

The magnetic field $B\leq 7$\,T has been applied along the c-axis
using a superconducting split-pair magnet equipped with mylar
windows for the electromagnetic radiation. The experiments in
magnetic field have been performed using the field sweeping (FS)
conditions \cite{matsuda}. Within the FS conditions the actual
field in the sample may deviate from the external field by the
value determined by the Bean critical field \cite{bean} $H_c =
J_c \delta /2$, where $J_c$ is the critical current density and
$\delta$ is the sample thickness. From the magnetization
measurements the Bean critical field has been determined as  $H_c
\simeq 500\,\textrm{Oe}$ at $T = 2$\,K, which corresponds to $J_c
\simeq 6.6\cdot 10^4\,\textrm{A/cm}^2$. The demagnetization
effects can be neglected because of the thin-plate geometry of
the experiment.

\begin{figure}[]
\centering
\includegraphics[width=7cm,clip]{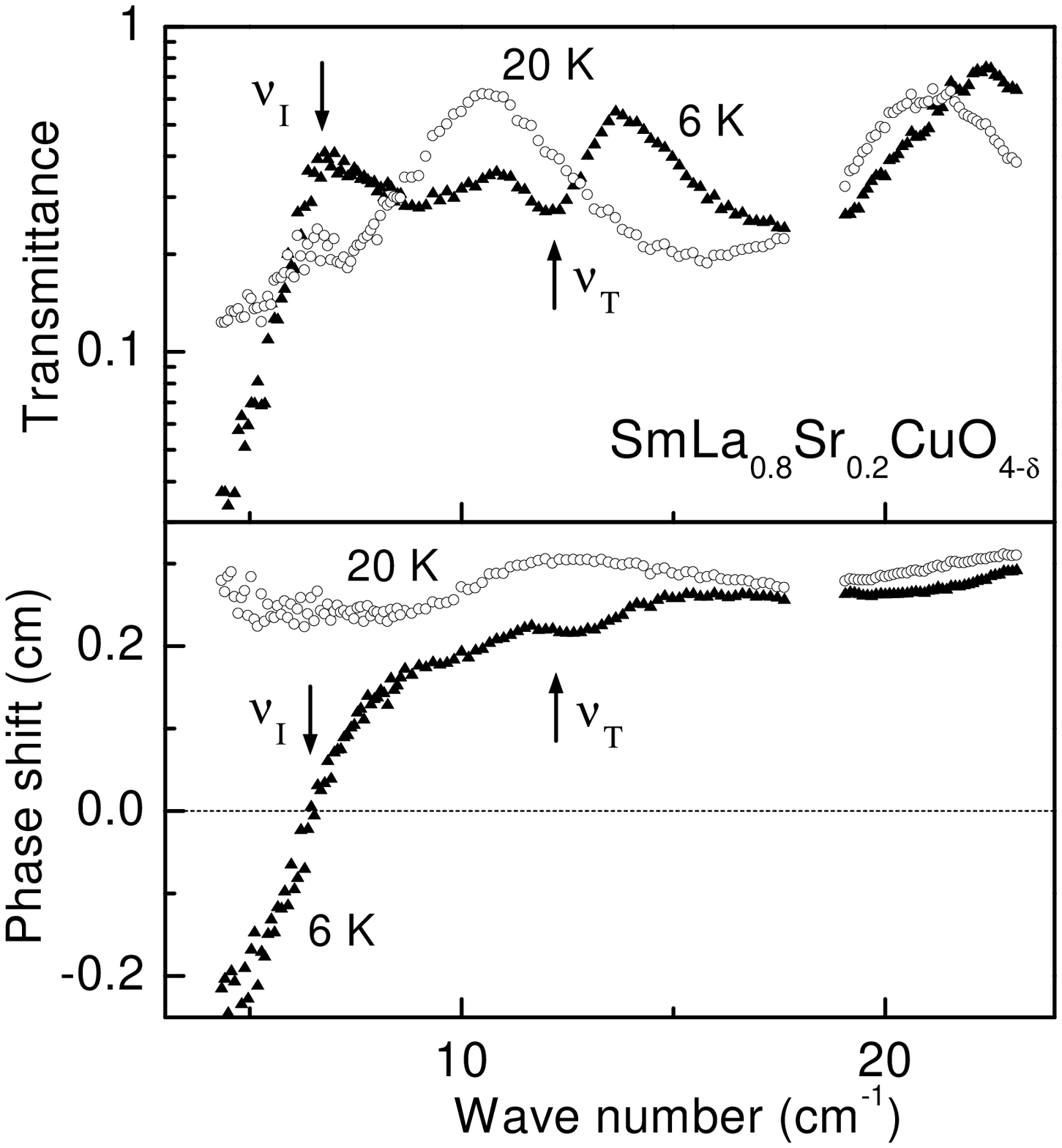}
\vspace{0.2cm} \caption{Transmittance and phase shift spectra of
\sm \ along the c-axis in zero magnetic field. Arrows indicate
the positions of the transverse ($\nu_{T}$) and of the low-lying
longitudinal plasmons($\nu_{K}$). } \label{transm}
\end{figure}

Fig.\ \ref{transm} shows  examples of  transmittance and phase
shift spectra of SmLSCO along the c-axis above and below the
superconducting transition. The overall frequency dependence of
the data is dominated by the interference fringes due to relative
transparency of the sample along the c-axis. In the
normally-conducting state ($T=20$\,K) the interference maxima are
seen around $\nu _{1}=10.6$\,cm$^{-1}$ and
$\nu_{2}=21.3$\,cm$^{-1}$, from which the c-axis dielectric
constant $\varepsilon_{c}=15.2$ can directly be estimated via $2
\delta \varepsilon _{c}^{2} = m/\nu _{\max }$. Here $m=1,2$ is an
integer. The measured curves in Fig.\ \ref{transm} are strongly
modified in the superconducting state: both transmittance and
phase shift are suppressed at low frequencies due to the growth
of the imaginary part of the complex conductivity $\sigma
^{*}=\sigma _{1}+i\sigma _{2}$. Most importantly, two new effects
can directly be observed in the experimental transmittance data
as an additional maximum at $\nu _{I}\simeq 6.7$\,cm$^{-1}$ and
minimum at $\nu _{T}\simeq 12.1$\,cm$^{-1}$. The first feature
corresponds to the (low-lying) longitudinal plasmon and is seen
also as a zero crossing of the phase shift. The second feature
corresponds to a transverse plasmon and is characterized by a
minimum in transmittance and a smooth step in the phase shift.
The second longitudinal plasmon ($\nu _{K}\simeq \nu
_{T}+0.2$\,cm$^{-1})$ is not seen in the transmittance spectra
because it strongly overlaps with the transverse resonance.

\begin{figure}[]
\centering
\includegraphics[width=7cm,clip]{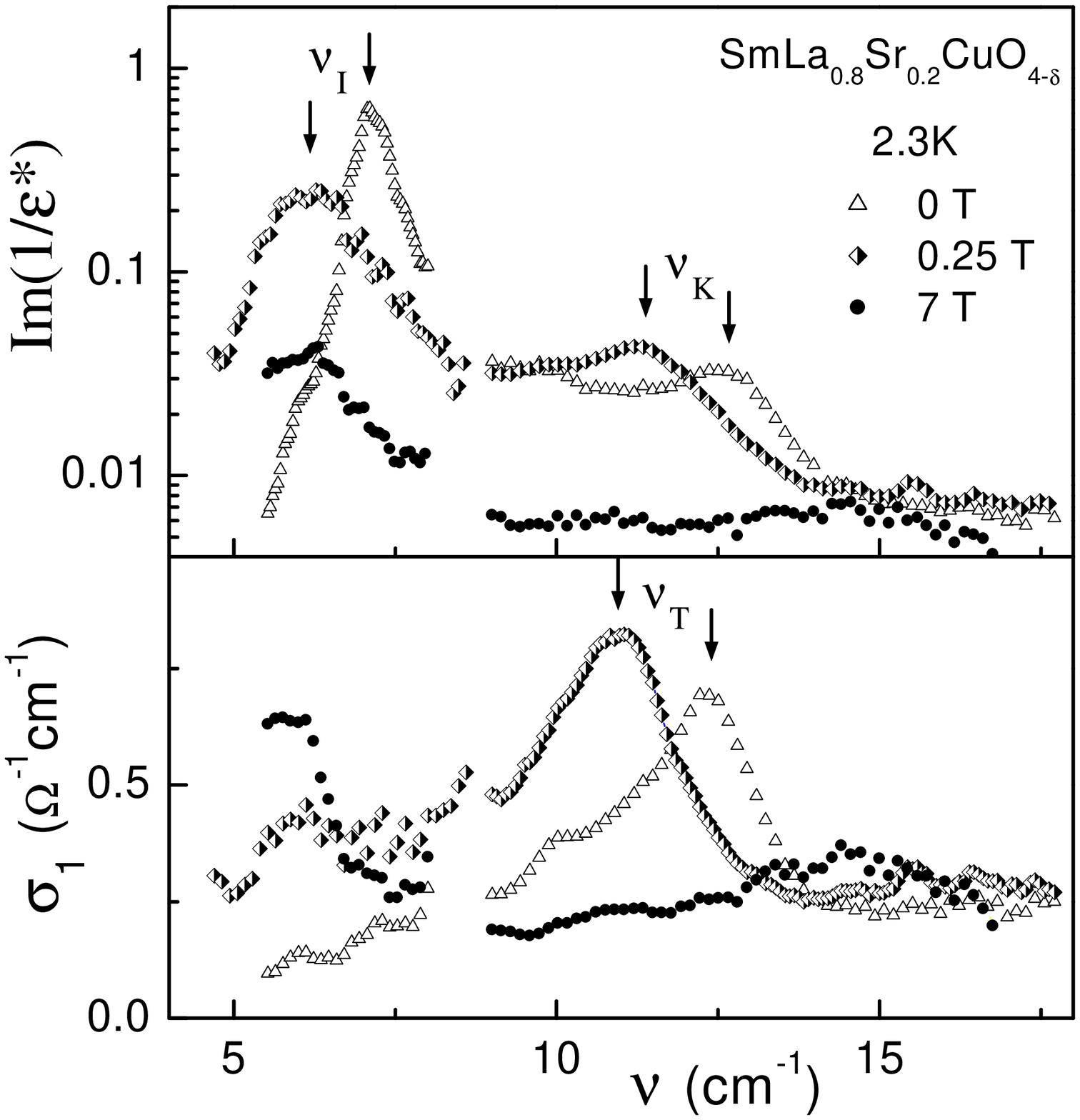}
\vspace{1cm} \caption{Real part of the complex conductivity
$\sigma_1$ and loss function $Im(1/\varepsilon^*)$ of \sm \ along
the c-axis for different magnetic fields at 2.3\,K. The transverse
plasmon $\nu_{T}$ shows up as peak in $\sigma_1$, the longitudinal
plasmons $\nu_{I,K}$ as peaks in $Im(1/\varepsilon^*)$.}
\label{cond}
\end{figure}

The evaluation of the complex conductivity from the transmittance
and phase shift  removes the interference fringes from the spectra
because they are automatically included in the Fresnel
expressions. The resulting conductivity $(\sigma _{1})$ and the
loss function $( \mathop{\rm Im} [1/\varepsilon ^{*}])$ are
represented in Fig.\ \ref{cond} for different magnetic fields
 along the c-axis. The complex dielectric function is given by
 $\varepsilon ^{*}=\varepsilon
_{1}+ i \varepsilon _{2}=i\sigma ^{*}/\varepsilon _{0}\omega $,
where $\varepsilon _{0}$ is the permittivity of free space and
$\omega =2\pi \nu $ is the angular frequency. The transverse
plasmon is clearly seen in the lower frame of Fig.\ \ref{cond} as
a maximum of $\sigma _{1}$. The longitudinal plasmons cannot be
observed in the real part of the conductivity because they do not
lead to the absorption of the electromagnetic radiation. Instead,
they can be observed as peaks in the loss function.

\begin{figure}[]
\centering
\includegraphics[width=7cm,clip]{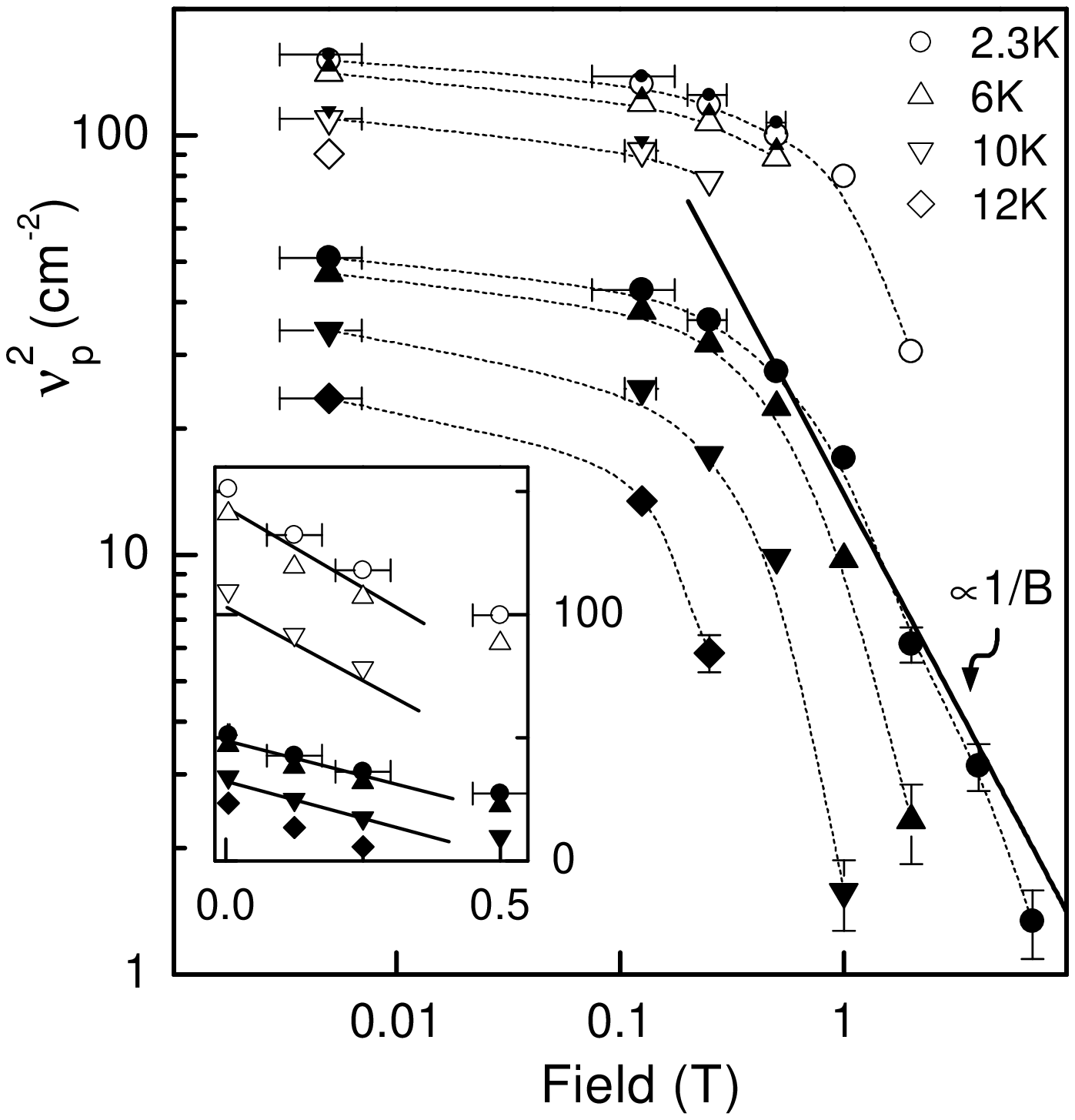}
\vspace{0.2cm} \caption{Magnetic field dependence of the (squared)
c-axis plasmons in \sm: Open symbols - transverse plasmon, large
and small closed symbols - longitudinal plasmons. Thin dashed
lines are guides to the eye. The thick solid line indicates a
$1/B$ field dependence. The inset represents the data on the
linear scale.} \label{freq}
\end{figure}

Fig.\ \ref{freq} represents the magnetic field dependence of the
plasma resonances in SmLSCO. The frequency of the transverse
plasmon has been determined by the peak positions in $\sigma _{1}$
and the frequency of the longitudinal plasmons have been derived
from the maxima of the loss function. For high fields and high
temperatures the low-lying longitudinal plasmon was shifted to low
frequencies, outside the range of the present experiment. In that
case, the resonance frequency has been obtained as zero crossing
of $\varepsilon _{1}$, assuming that in the superconducting state
the usual low-frequency relation holds: $\varepsilon _{1}\propto
-1/\nu ^{2}$.

For all resonances presented in Fig.\ \ref{freq} the
field-dependence can be clearly separated in two regimes: a
low-field region with a weak field dependence below $0.1-1.0$\,T
and a high-field regime, for which an approximate dependence $\nu
_{p}^{2}\propto 1/B$ is applicable. We identify therefore the
high field region as a vortex-liquid state, in which both,
theoretically \cite{koshelev} and experimentally
\cite{matsuda,shibaushi,gaifullin}, the plasma frequency has been
shown to behave as $\nu _{p}^{2}(B,T) / \nu _{p}^{2}(0)\propto
B^{-1}T^{-1}$.

In addition, the inset of Fig.\ \ref{freq} represents the
low-field dependence of the plasma frequencies on the linear
scale. From this presentation the linear field-dependence of $\nu
_{p}^{2}$ in low magnetic field becomes evident. At low fields
the regime of isolated vortices is expected to be applicable and
no field dependence of the plasma frequency is expected in zeroth
order approximation. However, taking into account thermal
fluctuation and pinning disorder, a linear correction of the
squared plasma frequency is expected theoretically
\cite{bulaevski} and has been recently observed \cite{dulic1} in
Tl$_{2}$Ba$_{2}$CaCu$_{2}$O$_{8}$ using the JPR technique.
However, we note that the theoretical considerations have been
carried out for the system with a single Josephson coupling
between the CuO$_2$ layers. In the zeroth order approximation, we
expect the same magnetic field dependencies also for a system
with two coupling constants like SmLSCO.

\begin{figure}[]
\centering
\includegraphics[width=7cm,clip]{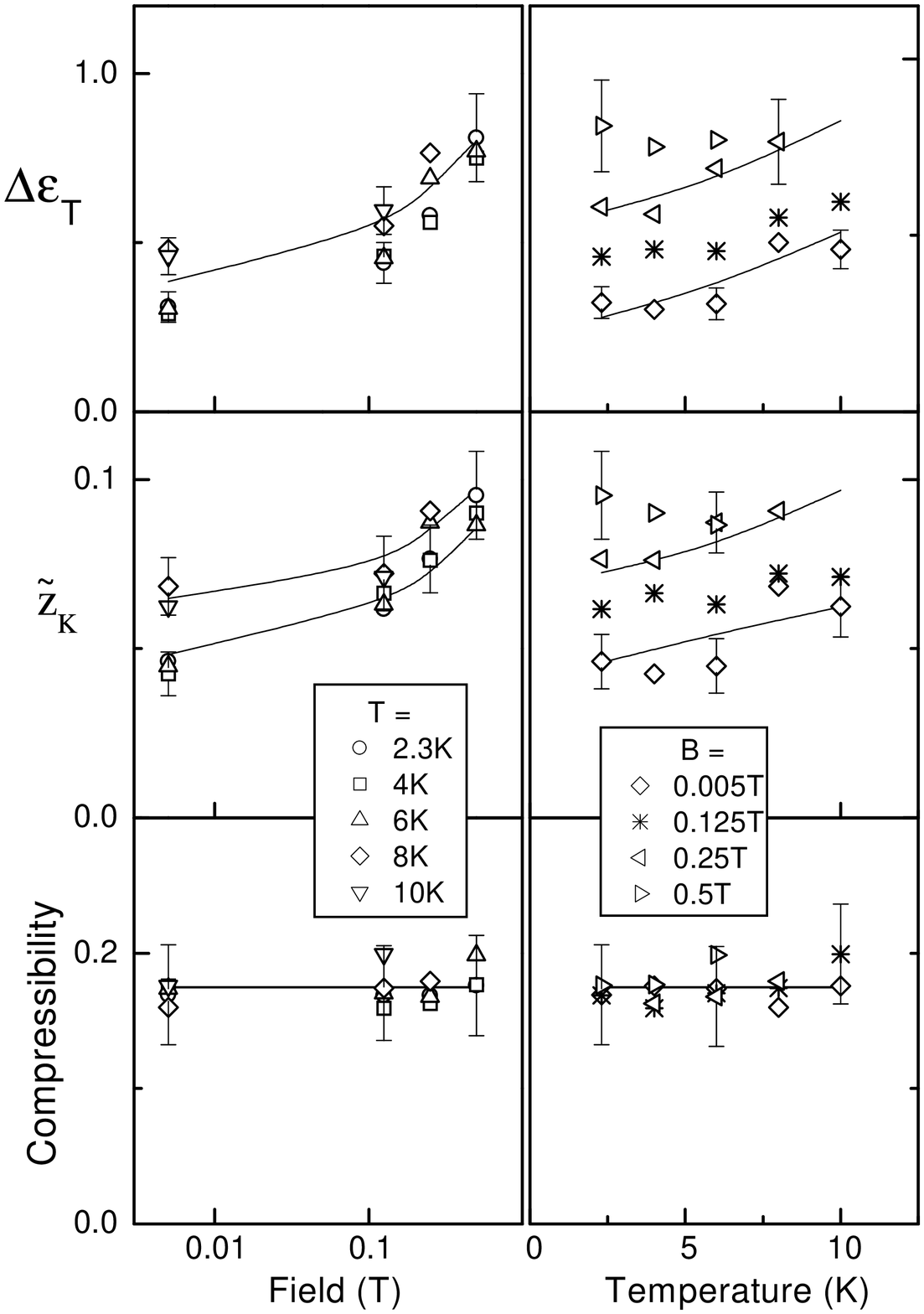}
\vspace{0.2cm} \caption{ Upper panels: magnetic field (left) and
temperature (right) dependence of the dielectric contribution of
the transverse plasmon in \sm. Middle panels: field (left) and
temperature (right) dependence of the weight factor
$\tilde{z}_K=1-\tilde{z}_I$ in SmLSCO, obtained according to the
multilayer model \protect \cite{marel1,marel2}. Lower panels:
electronic compressibility $\gamma
=\varepsilon_{0}\varepsilon_{\infty}/(de^{2}Kn^{2})$ in SmLSCO.
Lines are guides to the eye. } \label{comp}
\end{figure}

The upper panels of Fig.\ \ref{comp} represents the dielectric
contribution of the transverse plasmon $\Delta
\varepsilon_{T}=\omega_T^2 / \omega_p^2$, which may be obtained
either integrating the area under the resonance in $\sigma _{1}$
via $\omega_p^2 = \frac{2}{\pi\varepsilon_0}\int \sigma_1
(\omega) d\omega$ (Fig.\ \ref{cond}) or fitting a conventional
oscillator model to the complex conductivity. Both procedures led
to similar results. A reliable analysis of the spectra could be
carried out for $B\leq0.5\,$T only, which explains somewhat
limited range of data in Fig.\ \ref{comp}. As demonstrated by the
left upper panel of Fig.\ \ref{comp}, the magnetic field strongly
influences the dielectric contribution of the transverse plasma:
$\Delta\varepsilon_T$ increases by a factor of 2 between $B\simeq
0$ and $B=0.5$\,T. The field enhancement of $\Delta
\varepsilon_{T}$ leads to the non-conservation of the transverse
spectral weight $\omega_p^2=\Delta \varepsilon_{T}\cdot
\omega_{T}^{2}.$ According to Fig.\ \ref{freq}, $\omega_{T}^{2}$
is reduced by a factor of 1.5 at $B=0.5$\,T, which does not
compensate the doubling of the dielectric contribution and
indicates a $\sim 30\%$ enhancement of the transverse spectral
weight in magnetic field of 0.5\,T. The temperature dependence of
the dielectric contribution $\Delta \varepsilon _{T}(T)$ is shown
in the right upper panel of Fig.\ \ref{comp}. As in the case of
the field-dependence, the data indicate a weak increase with
temperature.

Finally, we discuss the results within the frame of the
multilayer model by van der Marel and Tsvetkov \cite{marel2}. In
this model the complex dielectric constant of a layer with two
different coupling constants is given by
\begin{equation}
\frac{\varepsilon _{\infty }}{\varepsilon ^{*}(\nu
)}=\frac{\tilde{z}_{I}\nu ^{2}}{\nu ^{2}-\nu _{I}^{2}+i\nu
g_{I}}+\frac{\tilde{z}_{K}\nu ^{2}}{\nu ^{2}-\nu _{K}^{2}+i\nu
g_{K}} \label{eeps}
\end{equation}
where $\nu _{I}^{2}$ and $\nu _{K}^{2}$ are low- and
high-frequency longitudinal plasma frequencies, $g_{I}$ and
$g_{K}$ are the corresponding damping factors, and $\varepsilon
_{\infty }\simeq 16$ is the high-frequency dielectric constant.
The transverse plasma frequency $\nu_{T}$ is given by
$\nu_{T}^{2}=\tilde{z}_{K}\cdot \nu_{I}^{2}+\tilde{z}_{I}\cdot \nu
_{K}^{2}$ . Here $\tilde{z}_{K}$ and $\tilde{z}_{I}$ are the
weight factors of both longitudinal plasmons, which are connected
as $\tilde{z}_{K}+\tilde{z}_{I}=1.$ The weight factors are
obtained via \cite{marel2}
\begin{equation}
\begin{array}{c}
\tilde{z}_{K}=\frac{1}{2}-\frac{2\gamma (z_{K}z_{I}+\gamma
)}{ z_{K}z_{I}+2\gamma +4\gamma
^{2}}\frac{\nu_{K}^{2}+\nu_{I}^{2}}{\nu_{K}^{2}-\nu_{I}^{2}}+ \\

\frac{ (z_{K}-z_{I})(z_{K}z_{I}+2\gamma )}{2(z_{K}z_{I}+2\gamma
+4\gamma ^{2})} \sqrt{1-\frac{4(2\gamma
)^{2}\nu_{K}^{2}\nu_{I}^{2} } {(\nu_{K}^{2}-\nu_{I}^{2})^{2}}}\;.
\end{array}
\label{ezk}
\end{equation}
Here $z_{K,I}$ are the unrenormalized weight factors which are
directly obtained from the relative distances $d_{1,2}$ between
the corresponding layers in SmLSCO
$z_{K,I}=d_{1,2}/(d_{1}+d_{2})$. This leads to $z_{K}\simeq
z_{I}\simeq 0.5$, because $d_{1}\simeq d_{2}$. A correction due
to the lattice polarizability, characterized by the dielectric
constants of the layers, gives \cite{dulic,marel2}
$z_{K}=1-z_{I}\simeq 0.43\pm 0.08$.

Despite its relative complexity, the advantage of Eq. (\ref{ezk})
is that it contains only one unknown parameter $\gamma $ which is
inversely proportional to the two-dimensional electronic
compressibility ($K$), $\gamma =\varepsilon _{0}\varepsilon
_{\infty }/(de^{2}Kn^{2})$. Here $n$ is the electron density and
$d=12.56\,$\AA \ is the lattice constant along the c-axis. The
frequencies $\nu_{K}^{2},\nu_{I}^{2}$ and the weight factor $
\tilde{z}_{K}$ are measured in the experiment, while the starting
weight factors $z_{K,I}$ are obtained from structural
considerations.

The data, presented in Fig.\ \ref{freq} and in the upper panel of
Fig.\ \ref{comp}, are already sufficient to determine all
parameters of the model, because the dielectric contribution of
the transversal plasma $\Delta \varepsilon_{T}$ can be obtained
from Eq. (\ref{eeps}) and is given by
\begin{equation}
\Delta \varepsilon _{T} = \varepsilon _{\infty
}\tilde{z}_{K}\tilde{z} _{I}(\nu_{K}^{2} - \nu_{I}^{2})^{2} /
\nu_{T}^{4}\;.  \label{ede}
\end{equation}
Eq.\ (\ref{ede}) provide the values of the weight factors from
known resonant frequencies and the dielectric contribution.
However, equivalent results were obtained by directly fitting the
experimental spectra using Eq.\ (\ref{eeps}).

The middle panels of Fig.\ \ref{comp} show the weight factor of
the high-frequency plasmon in SmLSCO as a function of temperature
and magnetic field. The weight factor $\tilde{z}_{K}$ reveal the
dependencies, which are similar to that of $\Delta \varepsilon
_{T}$ in the upper panel: a pronounced increase in magnetic field
and much weaker but still visible increase as function of
temperature.

A striking result of this work is shown in the lower panels of
Fig.\ \ref {comp}, which represent the electronic compressibility
of SmLSCO. The compressibility reveals neither field nor
temperature dependence in the presented range. As a matter of
fact, in first approximation it can be expected that the
compressibility $Kn^{2}=\partial n/\partial \mu $, where $\mu$ is
the chemical potential, is independent of field and temperature.
This strongly supports the idea that $\gamma $  indeed is the
reason for the deviation of the the simple version of the
multilayer model \cite{marel1} from experimental data
\cite{shibata2,kakeshita,dulic}. From the data of Fig.\
\ref{comp} we obtain $Kn^{2}=0.6\pm 0.2$\ $eV^{-1}$ per Cu-atom.
A discrepancy in absolute value to Ref. \cite{dulic}
($Kn^{2}=1.1$\ $eV^{-1}$) is due to the difference in the
high-frequency dielectric constant, which has been adopted for
calculations. We note further, that the electronic compressibility
of the two-dimensional electronic gas depends upon the effective
mass only $Kn^{2}=\partial n/\partial \mu =m_{eff}/(\pi \hbar
^{2})$ from which the effective mass may be estimated:
$m_{eff}=0.95m_{e}$.

In conclusion, we have measured the magnetic field and temperature
dependencies of the longitudinal and transverse plasmons in
SmLa$_{0.8}$Sr$ _{0.2}$CuO$_{4-\delta }$. In agreement with
previous data, two longitudinal and one transverse plasmon have
been observed in the submillimeter $(5<\nu <30$\,cm$^{-1})$
spectra. The field dependence of the resonance frequencies $\nu
_{p}^{2}$ reveals two different regimes. In the low-field region
$B<0.1-1.0$\,, which we identify as a vortex-glass regime, $\nu
_{p}^{2}$ linearly depends  upon field. For higher magnetic
fields, $\nu _{p}^{2}$ are much more strongly suppressed, which
may be approximated by $\nu _{p}^{2}\propto B^{-1.25}$ and is
close to $\nu _{p}^{2}\propto 1/B$ known from the theory of the
vortex-liquid state. The full experimental data-set was
successfully described by the multilayer model by van der Marel
and Tsvetkov \cite{marel2}, which includes the electronic
compressibility to correctly obtain the frequency and the
spectral weight of the transverse plasmon. The most surprising
result is the field and temperature {\em independence} of the
compressibility, with an absolute value close to that of a
two-dimensional noninteracting electron gas.

We would like to thank A. A. Tsvetkov and Ch. Helm for useful
discussions. We are indebted to M. M\"{u}ller for carrying out
the SQUID experiments. This work was supported by the BMBF via the
contract 13N6917/0 - EKM.

\end{multicols}
\end{document}